\documentclass[twocolumn,showpacs,preprintnumbers,prc,amsmath,amssymb,floatfix]{revtex4}

\usepackage{graphicx}
\usepackage{dcolumn}
\usepackage{bm}

\newcommand{\beqa}{\begin{equation}}
\newcommand{\eeqa}{\end{equation}}

\begin{document}

\title{Phenomenology of $f_0(980)$ photoproduction on the proton at energies measured with the CLAS facility}

\author{M. L. L. da Silva$^1$ and M. V. T. Machado$^2$}
\affiliation{$^1$ Instituto de F\'{\i}sica e Matem\'atica, Universidade Federal de
Pelotas\\
Caixa Postal 354, CEP 96010-090, Pelotas, RS, Brazil\\
$^2$ High Energy Physics Phenomenology Group, GFPAE  IF-UFRGS \\
Caixa Postal 15051, CEP 91501-970, Porto Alegre, RS, Brazil}

\begin{abstract}
In this work we present the results of a theoretical analysis of the data on photoproduction of $f_0(980)$ meson in the laboratory
photon energy between 3.0 and 3.8 GeV. A comparison is done to the measurements performed by the CLAS collaboration at JLab accelerator
for the exclusive reaction $\gamma p\rightarrow pf_0(980)$. In the analysis the partial S-wave differential cross section is described
by a model based on  Regge approach with reggeized exchanges and distinct scenarios for the $f_0(980)\rightarrow V\gamma$ coupling are
considered. It is shown that such a process can provide information on the resonance structure and production mechanism.    
\end{abstract}


\maketitle

\section{Introduction}

The spectroscopy of the low mass scalar mesons, like the $f_0(980)$ one, is an exciting issue in hadronic physics and is still an
unsettled question. The topic is quite controversial as the mass spectrum ordering of low-lying scalar mesons disfavours the usual
quark-antiquark picture and a window is opened to a prolific investigation of the topic. For instance, in the past years the low mass
states $J^{PC}=0^{++}$ have been considered quark-antiquark mesons \cite{Marco1}, tetraquarks \cite{Marco2}, hadron molecules
\cite{Marco3}, glueballs and hybrids \cite{Marco4,Ochs}.  Such a conflicting interpretation comes from the fact the situation is
complex in low energies where quantitative predictions from QCD are challenging and rely mostly on numerical techniques of lattice QCD.
Nowadays, theoretical analysis consider also the gluonic degrees of freedom as the glueball resonance with no quarks having not exotic
quantum numbers and that  cannot be accommodated within quark-antiquark nonets \cite{Crede}. In this context, the photoproduction of
exotic mesons \cite{DK}  can be addressed using arguments based on vector meson dominance where the photon can behave like an $S=1$
quark-antiquark system. Therefore, such a system is more likely to couple to exotic quantum number hybrids. This sort of process  could
provide an alternative to the direct observation of the radiative decays at low energies. Along those lines, the GlueX experiment
\cite{GlueX} is being installed and its primary purpose is to understand the nature of confinement in QCD by mapping the spectrum of
exotic mesons generated by the excitation of the gluonic field binding the quarks. 

The gluonic content of mesons could be directly tested in current accelerators in case a clear environment be available. In the limit
of very high energies the exclusive exotic meson production in two-photon and Pomeron-Pomeron interactions in coherent nucleus-nucleus
collisions can be easily computed. In these cases, the photon flux scales as the square charge of the beam,
$Z^2$, and then the corresponding cross section is highly enhanced by a factor $\propto Z^4\approx 10^7$ for gold or lead nuclei at RHIC
and LHC, respectively. A competing channel, which produces similar final state configuration, is the central diffraction process that
is modeled in general by two-Pomeron interaction. For instance, in Ref. \cite{SMdiff} the cross sections for these two channels are
contrasted in the production of glueball candidates like the low-lying scalar mesons. The cross sections were sufficiently large for
experimental measurement and the event rates can be obtained using the beam luminosity \cite{upcs} for LHC. It produces  $5\cdot 10^5$
events for $f_0(1500)$ mesons in the two-photon channel whereas the integrated cross section for exclusive diffractive process is around
500 $\mu$b.  The central diffractive production of mesons $f_0(980)$ and $f_2(1270)$ at the energy of CERN-LHC experiment on
proton-proton collisions was further investigated in Ref. \cite{MVTMALICE}. The processes initiated by quasi-real photon-photon
collisions and by central diffraction processes were also considered. The main motivation is that  ALICE collaboration has recorded zero
bias and minimum bias data in proton-proton collisions at a center-of-mass energy of $\sqrt{s}=7$ TeV. Among the relevant events, those
containing double gap topology have been studied and they are associated to central diffractive processes \cite{ALICE}. In particular,
central meson production was observed. In the double gap distribution, the $K_s^0$ and $\rho^0$ are highly suppressed while the
$f_0(980)$ and $f_2(1270)$ with quantum numbers $J^{PC}=(0,2)^{++}$ are much enhanced. Such a measurement of those states is evidence that
the double gap condition used by ALICE selects events dominated by double Pomeron exchange. The main predictions of Ref. \cite{MVTMALICE}
are the exclusive diffractive cross section for $f_0(980)$ being $d\sigma (y=0)/dy\simeq 27$ $\mu$b and its production cross section in two-photon reactions giving $\sigma_{\gamma\gamma}=0.12$ nb at $\sqrt{s}_{pp}=7$ TeV.

As far the low energy regime is concerned, in Ref. \cite{SM} we studied the photoproduction of the $a_0(980)$, $f_0(1500)$ and $f_0(1710)$
resonances for photon energies relevant for the GlueX experiment at $E_{\gamma}=9$ GeV using current ideas on glueball and $q\bar{q}$
mixing. The meson differential and integrated cross sections were evaluated and the effect of distinct mixing scenarios were investigated.
Although large backgrounds were expected, the signals could be visible by considering only the all-neutral channels, that is their decays
on $\pi^0\pi^0$, $\eta^0\eta^0$ and $4\pi^0$. The theoretical uncertainties were still large, with $f_0(1500)$ being the more optimistic case.  

The $f_0(980)$ photoproduction is measured via the most sizable decay modes which are $\pi\pi$ and $K\bar{K}$. In this mass range was
expected an interference of the P-wave from the decay of $\phi(1020)$ and the S-wave from decay of $f_0(980)$ ressonance. This interference
is discussed in Refs. \cite{SZC,SP} where the $\pi\pi$ and $K\bar{K}$ photoproduction was performed close to the $K\bar{K}$ threshold. The
photoproduction of $f_0(980)$ was investigated in a unitary chiral model \cite{oset}. Here, we investigate the photoproduction of meson
state $f_0(980)$ and distinct scenarios for the $f_0(980)\rightarrow V\gamma$ coupling are considered. The scenarios discussed in this
paper consider the $f_0(980)$ as a tetraquark or as a ground state nonet. We focus on the S-wave analysis on the forward photoproduction of
$\pi^+\pi^-$ on the final state. The theoretical formalism employed is the Regge approach with reggeized $\rho$ and $\omega$ exchange
\cite{SM}. This assumption follows from Regge phenomenology where high-energy amplitudes are driven by $t$-channel meson exchange. This
paper is organized as follows: in next  section we present the relevant scattering amplitudes and how they are related to the differential
cross sections.  In last section numerical results are presented and parameter dependence is addressed. A comparison to the CLAS data for
direct photoproduction of $f_0(980)$ is also done \cite{CLASf0}. Finally, the conclusions and discussions follow at the end of section.

\section{Model and cross section calculation}

We focus on the S-wave analysis of nondiffractive $f_0(980)$ photoproduction and its decay on $\pi^+\pi^-$ final state. According to the Regge phenomenology,
one expects that only the $t$-channel meson exchanges are important in such a case. The $\rho$ and $\omega$ reggeized exchanges are to be
considered in the present analysis. To obtain mass distribution for the scalar $f_0(980)$ meson, one represents it  as relativistic
Breit-Wigner resonance with energy-dependent partial width. The differential cross section for the production of a scalar with invariant mass
$M$, and its decay to two pseudo-scalars, masses $m_a$ and $m_b$, can be written as,
\begin{eqnarray}
\frac{d\sigma}{dt~dM}=\frac{d\hat{\sigma}(t,m_S)}{dt}\frac{2m_S^2}{\pi}
\frac{\Gamma_i(M)}{(m_S^2-M^2)^2+ (M\Gamma_{\rm Tot})^2} ,
\label{signaldcs}
\end{eqnarray}
where $d\hat{\sigma}/dt$ is the narrow-width differential cross section at a scalar mass $M=m_S$ and $\Gamma_i(M)$ being the
pseudoscalar-pseudoscalar final state  partial width, which can be computed in terms of the $SPP$ coupling $g_i$. A note is in order
at this point. Although the main decay of the $f_0(980)$ is $\pi\pi$,  this state resides at the $K \bar K$ threshold. Therefore,
following Ref. \cite{DK} we use the Breit-Wigner parametrisations obtained in the analysis of $\phi$ radiative decays \cite{KLOE07f}. In
such a case, the Breit-Wigner width takes the form
\begin{eqnarray}
\Gamma(M) \!\! & = & \!\!  \frac{g_{\pi\pi}^2}{8\pi M^2}\sqrt{\frac{M^2}{4}-M^2_{\pi\pi}} \nonumber \\
\!\!\!\! \!\! \!\!  & + &\!\!  \frac{g_{K\bar{K}}^2}{8\pi M^2}\!\left[\sqrt{\frac{M^2}{4}-
M^2_{K^+K^-}} + \sqrt{\frac{M^2}{4}-
M^2_{K^0\bar{K}^0}}  \right],\nonumber
\\ & & 
\label{width_s}
\end{eqnarray}
where we set the following parameters: $M=984.7$ MeV, $g_{K\bar{K}}\equiv g_{K^+K^-}=g_{K^0 \bar K^0}=0.4$ GeV,
$g_{\pi^+\pi^-}=\sqrt{2}g_{\pi^0\pi^0}=1.31$ GeV for the scalar meson $f_0(980)$ considered here. However, when we consider the
$f_0(980)$ cross section below $K\bar{K}$ threshold the total width cannot be written as in Eq. (\ref{width_s}). Thus, in this case we should use the Flatt\`e formula \cite{flatte} when computing our numerical results in next section.

Let us proceeding, the reaction proposed here is $\gamma p \rightarrow p\,f_0(980)$.  Within the Regge phenomenology the differential
cross section in the narrow-width limit for a meson of mass $m_S$ is given by \cite{SM},
\begin{eqnarray}
\frac{d\hat{\sigma}}{dt}(\gamma p \rightarrow p M)= \frac{|{\cal M}(s,t)|^2}{64\pi\,(s-m_p^2)^2},
\label{dsigma}
\end{eqnarray}
where ${\cal M}$ is the scattering amplitude for the process, $s,\,t$ are usual Mandelstan variables and $m_p$ is the proton mass.
For the exchange of a single vector meson, i.e. $\rho$ or $\omega$ one has:
\begin{eqnarray}
|{\cal M} (s,t)|^2 &=& -\frac{1}{2}{\cal A}^2(s,t)\left[\frac{}{}s(t-t_1)(t-t_2) \right.\nonumber\\
      & & \,\,\,\,\,\,\,\, \left.    +\frac{1}{2}t(t^2 - 2 (m_S^2 +s)t + m_S^4) \right] \nonumber\\
& - &{\cal A}(s,t){\cal B}(s,t)m_ps(t-t_1)(t-t_2) \nonumber\\
& - &\frac{1}{8}{\cal B}^2(s,t)s(4m_p^2-t)(t-t_1)(t-t_2).
\label{msquare}
\end{eqnarray}
where $t_1$ and $t_2$ are the kinematical boundaries
\begin{eqnarray}
\!\!\!\!\!\!t_{1,2} &=& \frac{1}{2s}\left[-(m_p^2-s)^2+m_S^2(m_p^2+s)\right. \nonumber\\
& \pm & \left. (m_p^2-s)\sqrt{(m_p^2-s)^2-2m_s^2(m_p^2+s)+m_S^4}\, \right],
\label{t12}
\end{eqnarray}
and where one uses the standard prescription for Reggeising the Feynman propagators  assuming a linear Regge trajectory
$\alpha_V(t)= \alpha_{V0}+ \alpha^\prime_V t$ for writing down the quantities ${\cal A}(s,t)$ and ${\cal B}(s,t)$:
\begin{eqnarray}
{\cal A}(s,t) & = & g_A\,\left(\frac{s}{s_0}\right)^{\alpha_V(t)-1}\frac{\pi\alpha^\prime_V}{\sin(\pi\alpha_V(t))}
\frac{1-e^{-i\pi\alpha_V(t)}}{2\,\Gamma(\alpha_V(t))},\nonumber \\
{\cal B}(s,t) & = & -\frac{g_B}{g_A}\,{\cal A}(s,t).
\label{abdef}
\end{eqnarray}
It is assumed non-degenerate $\rho$ and $\omega$ trajectories $\alpha_{V}(t) = \alpha_{V}(0)+\alpha^{\prime}_{V}t$, with
$\alpha_{V}(0)=0.55\,(0.44)$ and $\alpha^{\prime}_{V} = 0.8\,(0.9)$ for $\rho$ ($\omega$). In Eq. (\ref{abdef}) above, one has that
$g_A = g_S(g_V+2 m_p g_T)$ and $g_B=2g_Sg_T$. The quantities $g_V$ and $g_T$ are the $VNN$ vector and tensor couplings, $g_S$ is the
$\gamma V N$ coupling. For the  $\omega N N$ couplings  we have set $g_V^{\omega} = 15$ and $g_T^\omega =0$ \cite{SM} and for the
$\rho N N$ couplings we used $g_V^{\rho} = 3.4$, $g_T^{\rho}= 11$ GeV$^{-1}$. The $S V \gamma$ coupling, $g_S$, can be obtained from
the radiative decay width through \cite{KKNHH}
\begin{eqnarray}
\Gamma(S \to \gamma V) = g_S^2\frac{m_S^3}{32\pi}\Bigg(1-\frac{m_V^2}{m_S^2}
\Bigg)^3.
\label{width}
\end{eqnarray}

This model was applied to $f_0(1370)$, $f_0(1500)$ and $f_0(1710)$ mesons which  are considered as mixing of $n\bar{n}$, $s\bar{s}$
and glueball states \cite{CDK}. In this case their radiative decays into a vector meson are expected to be highly sensitive to the
degree of mixing between the $q\bar{q}$ basis and the glueball. In Ref. \cite{SM} three distinct mixing scenarios were considered.
The first one is the bare glueball being lighter than the bare $n\bar{n}$ state; the second scenario corresponds to the glueball mass
being between the $n\bar{n}$ and $s\bar{s}$ bare state and finally the third one where glueball mass is heavier than the bare
$s\bar{s}$ state. The numerical values for the widths having effects of mixing on the radiative decays of the scalars on $\rho$ and
$\omega$ can be found in Table 1 of Ref. \cite{SM}. This way it is clear that the width is strongly model dependent and different
approaches can be taken into account. For instance, we quote the work in Ref. \cite{Jacosa}, where the decays of a light scalar meson
into a vector meson and a photon, $S\rightarrow V\gamma$, are evaluated in the tetraquark  and quarkonium assignments of the scalar
states. The coupling now reads,
\begin{eqnarray}
\Gamma(S \to \gamma V) = g_S^2\frac{(m_S^2 - m_V^2)^3}{8\,\pi\, m_S^3}.
\label{width_g}
\end{eqnarray}
The different nature of the couplings corresponds to distinct large-$N_c$ dominant interaction Lagrangians. In next section we compare
those approaches discussed above for the direct $f_0(980)$ photoproduction in the CLAS energies.

\begin{table}[t]
\begin{center}
\begin{tabular} {|c|c|}
\hline
Scenario & $f_0(980)\rightarrow \gamma V$  \\
\hline
\hline
1 & 83 (9.2)  \\
\hline
2 & 69880 (6730)  \\
\hline
3 & 3.3 (0.61)  \\
\hline
4 & 1005 (463)  \\
\hline
5 & 3.1 (3.4)  \\
\hline
\end{tabular}
\end{center}
\caption{\it The widths, $\Gamma(S\rightarrow \gamma V)$, for the radiative decays of the scalar meson into vector mesons 
$V=\rho\,(\omega)$  in units of keV. These results are extracted from Ref. \cite{DK} for scenario 1 and from Ref. \cite{Jacosa}
for the remaining scenarios.}
\label{tab1}
\end{table}

\section{Results and discussions}

In what follows we present the numerical results for the direct $f_0(980)$ photoproduction considered in present study and the
consequence of the tetraquark and quarkonium assignments of the scalar states discussed in previous section. The results
presented here will consider five distinct scenarios, three of them assuming that $f_0(980)$ is a quarkonium and two assuming
that $f_0(980)$ is a tetraquark. In scenarios 1, 2 and 3 the $f_0(980)$ will be interpreted as a ground-state nonet and in
scenarios 4 and 5 as a tetraquark. The $g_S$ coupling can be obtained from the radiative decay width in Table \ref{tab1}
using Eq. (\ref{width}) for scenario 1 and Eq. (\ref{width_g}) for the remaining scenarios. The values for scenario 1 in Table 
\ref{tab1} were extracted from Refs.~\cite{DK,KKNHH} and from Ref. \cite{Jacosa} for the remaining ones. The radiative decay in
scenarios 3 and 5 have considered the $f_0$ resonance as a quarkonium and a tetraquark, including Vector Meson Dominance, as discussed in Ref. \cite{Jacosa}.

\begin{figure}[ht]
\includegraphics[scale=0.34]{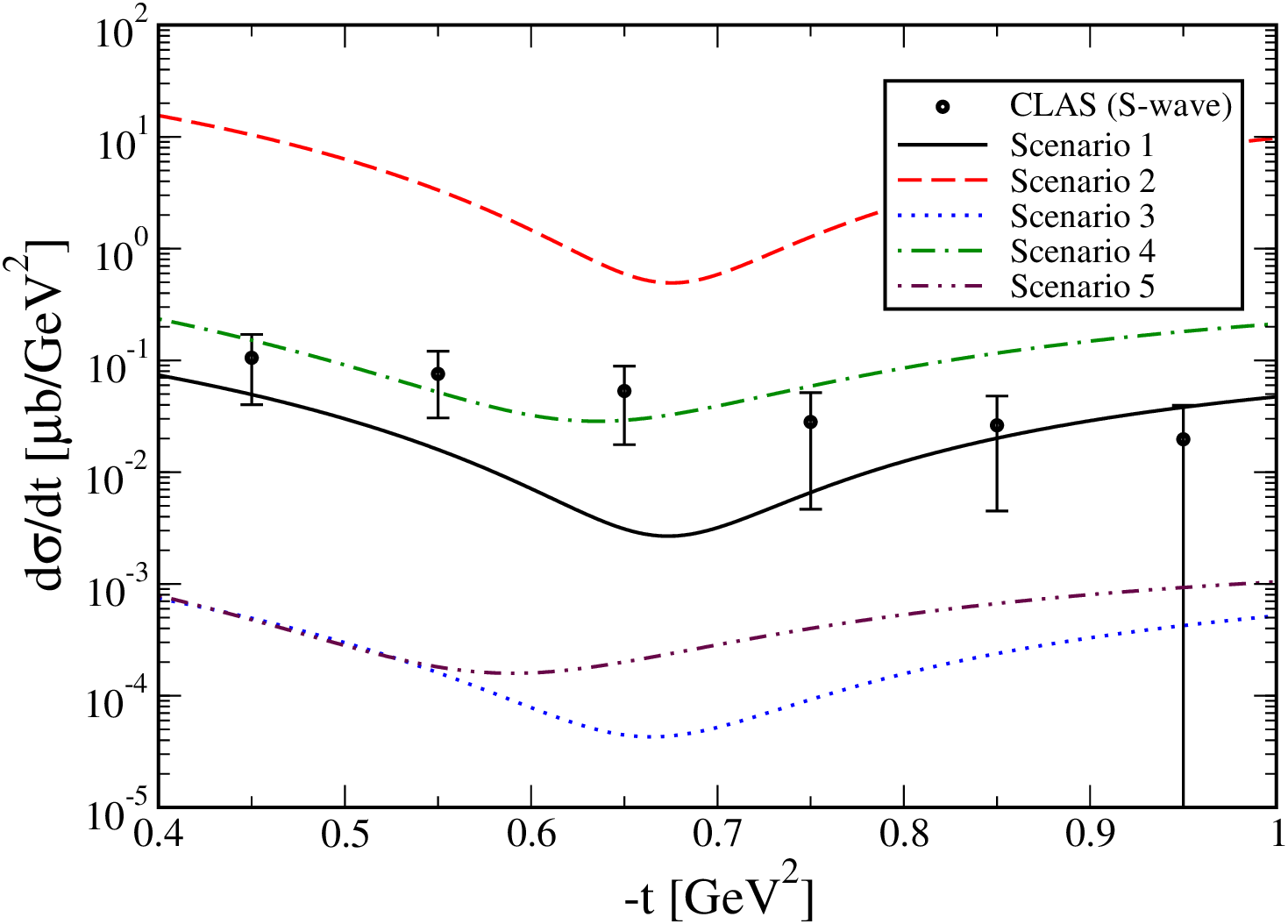}
\caption{(Color online) The $S$-wave differential photoproduction cross section  for $f_0(980)$ photoproduction as a function of
momentum transfer squared at CLAS experiment energy $E_{\gamma}=  3.4$ GeV. The statistical/systematic error bars for CLAS data
\cite{CLASf0}  were summed in quadrature.}
\label{fig:1}
\end{figure}

The partial S-wave differential cross sections for the $f_0(980)$ are presented in Fig. \ref{fig:1} at $E_{\gamma} = 3.4\pm 0.4$ GeV and
integrated in the $M_{\pi\pi}$ mass range $0.98\pm 0.04$ GeV. As a general picture, the typical pattern is a vanishing cross section
towards the forward direction (it does not appear in the plot as we are showing the region $|t|\geq 0.4 $ GeV$^2$) due to the helicity flip at the photon-scalar vertex and the dip at $ -t \approx 0.7$ GeV$^2$ related
to the reggeized meson exchange. The scenarios 1 and 4 are represented by the solid and dot-dashed lines, respectively. Both fairly reproduce the trend of CLAS data. The scenario 2 is denoted by the dashed curve. In this case the result overestimates the CLAS data points by a factor fifty. The scenarios 3 and 5 are represented by the dotted and dot-dot-dashed lines, respectively. Now, the results  underestimate the data by the same factor.

The several theoretical predictions were compared to the CLAS analysis at Jefferson Lab \cite{CLASf0}, where the $\pi^+\pi^-$ photoproduction
at photon energies between 3.0 and 3.8 GeV has been measured in the interval of momentum transfer squared $0.4 \leq |t|\leq 1.0$
GeV$^2$. There, the first analysis of S-wave photoproduction of pion pairs in the region of the $f_0(980)$ was performed. The
interference between $P$ and $S$ waves at $M_{\pi\pi}\approx 1$ GeV clearly indicated the presence of the $f_0$ resonance. As a final comment on the compatibility of theoretical predictions and experimental results, the scenarios 1 and 4 
 fairly describe the data (they are  parameter-free predictions as we did not fit any physical parameter).  Moreover, the CLAS data have no indication of a minimum as predicted by the reggeized models. Here, we have two possibilities: there is no data point in the dip region (around $|t|\simeq 0.7$ GeV$^2$) to confirm the reggeized exchange prescription or some additional contribution, i.e. background effects or interference, is missing. The case seems to be the the first option based on the reasonable description of $S$-wave by a non-reggeized meson exchange \cite{SP} as presented in Fig. 3 of Ref. \cite{CLASf0}.

As a complementary study, we also investigate the invariant mass distribution predicted by the theoretical models, taking the scenario 1 as a baseline. Another way to calculate the mass distribution is use the branching fractions for the strong decay of $f_0(980)$ associated with
the Breit-Wigner width for $f_0(980)$ to $\pi\pi$. In what follows two possibilities for the branching fractions will be used \cite{DK,lhcb}: 
\begin{eqnarray}
{\cal B} (f_0(980)\to \pi\pi) = 85 \% 
\label{b1}
\end{eqnarray}
and
\begin{eqnarray}
{\cal B} (f_0(980)\to \pi^+\pi^-) = 46 \pm 6 \% .
\label{b2}
\end{eqnarray}
On the other hand, it is possible to use the experimental value for the total width of $f_0(980)$ which is in the range of 40 to 100 MeV
\cite{pdg}. With this assumptions it is not necessary to calculate the $f_0(980) \to K\bar{K}$ width appearing in Eq. (\ref{width_s}).

\begin{figure}[ht]
\includegraphics[scale=0.32]{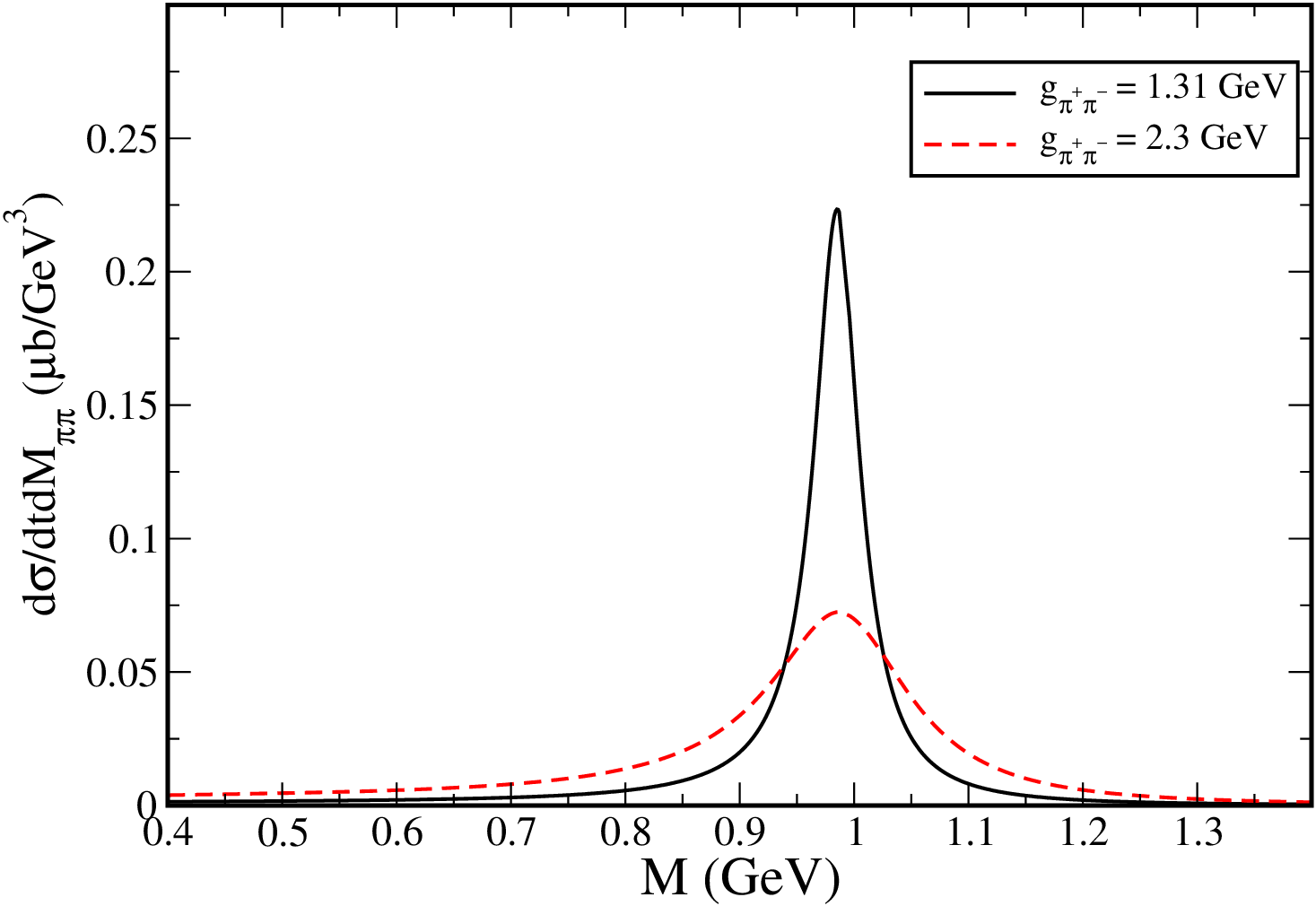}
\caption{(Color online) $S$-wave $\pi^+\pi^-$ invariant mass distribution at $E_{\gamma}=  3.4 $ GeV, $|t|=0.55$ GeV$^2$. The results
stand for $g_{\pi\pi} = 1.31$ GeV (solid curve) and $g_{\pi\pi} = 2.3$ GeV (dashed curve). In both cases, the value $g_{K\bar{K}} = 0.4$ is considered. }
\label{fig:2}
\end{figure}

In Fig. \ref{fig:2} the $S$-wave $\pi^+\pi^-$ invariant mass distribution at $E_{\gamma}=  3.4 $ GeV and $|t|=0.55$ GeV$^2$ is shown. For
the theoretical analysis we take the scenario 1.  One considers the coupling $g_{K\bar{K}} = 0.4$ and two possibilities for
the coupling $g_{\pi\pi}$. The first one is $g_{\pi\pi} = 1.31$ GeV (solid curve) presented in Ref. \cite{DK} and the second case $g_{\pi\pi} = 2.3 \pm 2$ GeV (dashed curve) given in Ref. \cite{coupling}. The Flatt\`e formula is used to obtain the $f_0(980)$ total width \cite{flatte}. The results present a strong dependence of the mass distribution on the $g_{\pi\pi}$ coupling.

\begin{figure}[ht]
\includegraphics[scale=0.32]{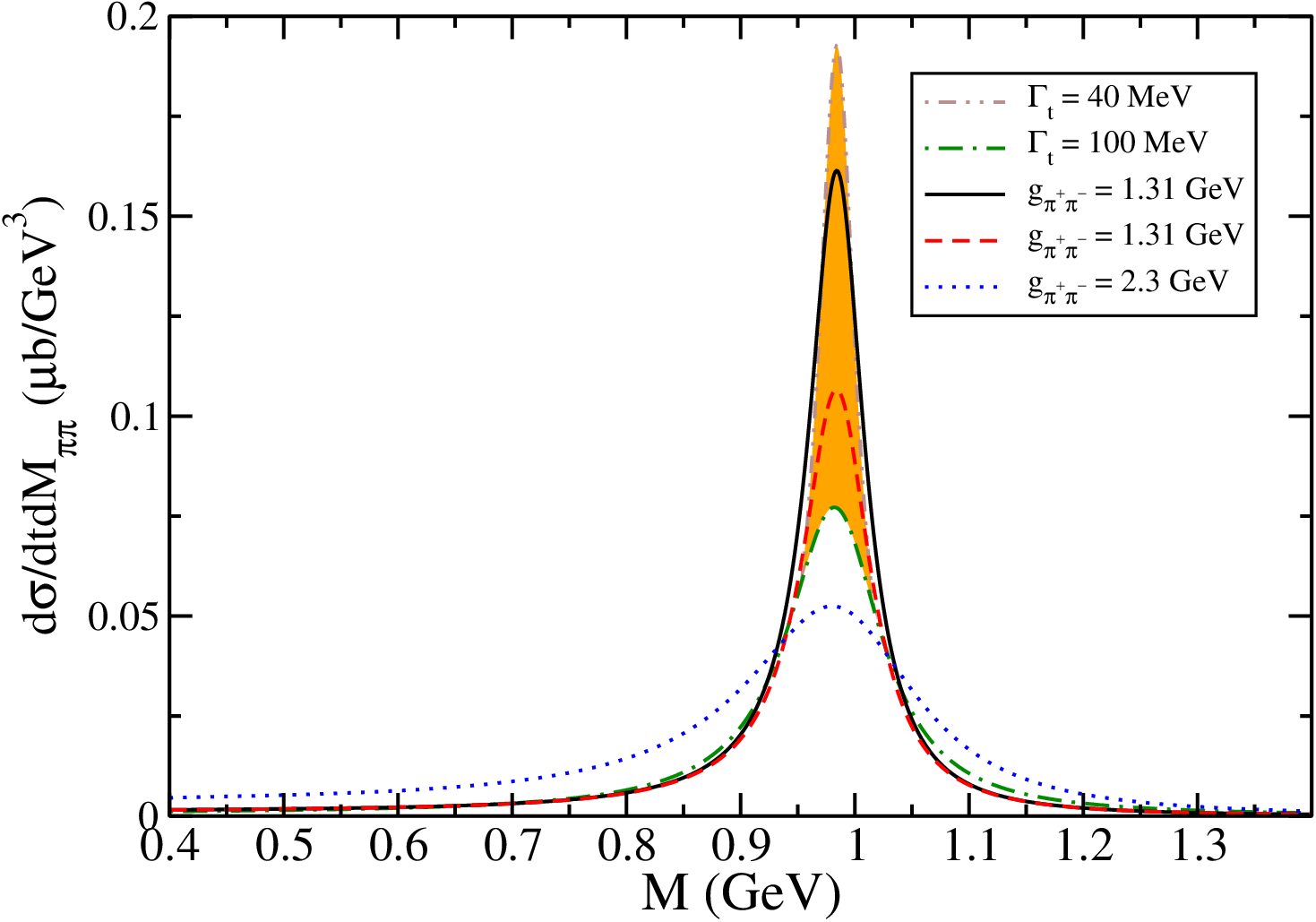}
\caption{(Color online) Analysis of the $S$-wave $\pi^+\pi^-$ invariant mass under several  consideration on the $f_0(980)$ partial width (see text). }
\label{fig:3}
\end{figure}

In Fig. \ref{fig:3} we repeat the previous analysis using now the experimental values $\Gamma_{tot}=40 - 100$ MeV for the $f_0(980)$ total width \cite{pdg}. It can be also obtained by branching ratios for $f_0(980)$ into pions associated with the Breit-Wigner formula. The dot-dot-dashed and dot-dashed lines represent the
invariant mass distribution for $\Gamma_{tot} = 40$ MeV  and $\Gamma_{tot} = 100$ MeV, respectively. In this case, $\Gamma_{\pi^+\pi^-} = 0.46 \Gamma_{tot}$ following Eq. (\ref{b2}). The solid  and dotted lines represent invariant mass distribution for $\Gamma_{tot} = \Gamma_{\pi\pi}/0.85$ following Eq. (\ref{b1}) and where $\Gamma_{\pi\pi}$ is given by Breit-Wigner formula. In the result represented by the dashed line $\Gamma_{tot} = \Gamma_{\pi^+\pi^-}/0.46$ following
Eq. (\ref{b1}) and $\Gamma_{\pi\pi}$ is given by Breit-Wigner formula. As indicated in  Fig. \ref{fig:2} there is a strong dependence on the coupling constant $g_{\pi\pi}$. An interesting dependence on branching ratios is observed too.

In summary, we have studied the photoproduction of $f_0(980)$ resonance for photon energies considered in the CLAS experiment at
Jefferson Lab, $E_{\gamma} = 3.4\pm 0.4$ GeV. It provides a test for current understanding of the nature of the scalar resonances.
We have calculated the differential cross sections as function of effective masses and momentum transfers. The effect of distinct
scenarios in the calculation of coupling  $S\rightarrow V\gamma$ was investigated. This study shows that we need to known more
precisely the radiative decay rates for $f_0(980)\to \gamma V$ which are important in the theoretical predictions. Our predictions of the cross
sections are somewhat consistent with the experimental analysis from CLAS Collaboration, at least for scenarios 1 and 4. The present experimental data are able to exclude
some possibilities for the  $S\rightarrow V\gamma$  coupling. We show also the large dependence on the model parameters as $g_{\pi\pi}$ value  and
branching fractions.

\begin{acknowledgments}
One of us (M. L. L. S.) would like to thanks M. L. Moreira for helpful discussions. This research was
supported by CNPq and FAPERGS, Brazil. 
\end{acknowledgments}

\end{document}